# Efficient Edge Detection on Low-Cost FPGAs

Jamie Schiel
Department of Mechatronic Engineering
University of Canterbury
Christchurch, New Zealand
jms314@uclive.ac.nz

Dr. Andrew Bainbridge-Smith
Department of Electrical and Computer Engineering
University of Canterbury
Christchurch, New Zealand
andrew.bainbridge-smith@canterbury.ac.nz

*Abstract*—Improving the efficiency of edge detection in embedded applications, such as UAV control, is critical for reducing system cost and power dissipation. Field programmable gate arrays (FPGA) are a good platform for making improvements because of their specialised internal structure. However, current FPGA edge detectors do not exploit this structure well. A new edge detection architecture is proposed that is better optimised for FPGAs. The basis of the architecture is the Sobel edge kernels that are shown to be the most suitable because of their separability and absence of multiplications. Edge intensities are calculated with a new 4:2 compressor that consists of two custom-designed 3:2 compressors. Addition speed is increased by breaking carry propagation chains with look-ahead logic. Testing of the design showed it gives a 28% increase in speed and 4.4% reduction in area over previous equivalent designs, which demonstrated that it will lower the cost of edge detection systems, dissipate less power and still maintain high-speed control.

## I. Introduction

Edge detection is a fundamental algorithm in image analysis that extracts important structural information. For many applications, structural information of an image is all that is needed for high-level functionality. Extracting this information reduces the amount of data to be processed and increases accuracy. Recent developments in low-cost, video-based stability control for unmanned aerial vehicles (UAV) have used edge detection to find the horizon line [1]. Before these developments, most stability control was based on optical flow algorithms. However, finding the horizon line with edge detection has proven to be just as effective and far less computationally intensive. This has been an important development for UAVs because it decreases system cost and power dissipation, which extends flight-time.

Flight control has to be performed on-the-fly, requiring on-board hardware that is efficiently designed to minimise power dissipation. Three commonly used hardware platforms for performing low-power edge detection are general-purpose processors (GPP), application-specific integrated circuits (ASIC) and field programmable gate arrays (FPGA). As image size and computational complexity of the edge detection algorithm increases, it becomes difficult to maintain real-time performance on GPPs. GPPs are often interfaced with digital signal processors (DSP) to increase the performance of certain tasks. DSPs and ASICs have highly specialised internal architectures that perform a limited set of tasks at high data-rates. However, this comes at the cost of re-programmability. FPGAs are a useful median between the re-programmability of microprocessors and specialised internal architecture of DSPs and ASICs. An FPGA allows digital circuits to be soft-wired in a reconfigurable environment. This high-level of design control allows greater control over the performance characteristics and makes FPGAs better suited to designing low-power edge detectors.

Many projects have implemented edge detection on FPGAs [2, 3, 4]. Most of them have been designed from a mathematical perspective to reduce the computational complexity of edge detection algorithms. This approach is very beneficial to the speed and area cost of ASIC and FPGA designs. In the case of FPGA design, however, there are more than just mathematical ways to reduce computational complexity that derive from the internal structure of a FPGA. This paper outlines a new edge detection architecture for FPGAs that significantly increases speed and area efficiency by leveraging mathematical properties of the algorithm and hardware properties of an FPGA. Advances in this area will reduce the cost and power dissipation of embedded edge detection systems, which is important for extending the capabilities of UAVs.

Section II presents some background on the problem in two parts: *1)* A discussion of the general mathematical process for finding edges in a video feed. From this information and its implications for design efficiency, an edge detection algorithm is recommended for low-cost FPGAs; *2)* Internal structure of an FPGA is summarised with a focus on power dissipation and the usefulness of its look-up tables and fast-carry chains for edge detection. The new FPGA architecture is outlined in Section III, which also describes the design implementation on an Altera FPGA. Section IV compares the new design to similar projects on GPPs, ASICs and FPGAs to verify the use of FPGAs for efficient, low-cost edge detection and test the performance of the new architecture. Success of the new design is determined by reduced system cost and power dissipation.

## II. Background

### A. Edge Detection

Edge detection is most commonly performed with a mathematical two-dimensional (2D) convolution. Edge intensity of an image pixel is the result of a dot-product of a convolution kernel with the overlapping area of the image (Fig.

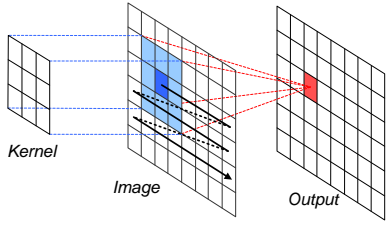

Fig. 1. Diagrammatic representation of a two-dimensional convolution showing how edge intensity output is produced from a dot-product of a convolutoin kernel and an image. Arrows on the "Image" show how the kernel slides over the image to cover all pixels.

1). In a physical sense, this calculation examines the intensity of pixels surrounding each point. Points where there is a rapid change in local intensity, which equates to a large dot-product, indicate the location of object boundaries and surface details.

There are many convolution kernels that detect edges with varying complexity and quality of results. In general, larger convolution kernels give better results because they measure variation over a larger area but are more computationally complex, which is ill-suited to low-cost FPGAs for reasons discussed in Section II-B. Convolution with an $n \times n$ size kernel requires $2n^2-1$ mathematical operations (multiplications and additions) in an FPGA. The Prewitt compass used by Juneja and Sandhuis [2] is the smallest edge detector with a size of $2 \times 2$ and therefore the least complex. Their results showed that the kernels are too small because image output suffered from large amounts of noise, where edges are falsely identified. The Canny detector, used in Neoh and Hazahchuk's [3] design, is often considered the best edge detector because of the high quality of results. However, the Canny detector requires several convolutions with large $5 \times 5$ kernels. Stowers et al. [1] showed that the level of precision produced by Canny detection is often not required. They demonstrated that the $3 \times 3$ bi-directional Sobel filters provide all the precision needed for accurate UAV control. The two Sobel filters, shown in (1) and (2), find edges in the horizontal and vertical directions. Bi-directional results are combined with a Euclidean two-norm like vectors.

$$Sx = \begin{bmatrix} -1 & 0 & 1 \\ -2 & 0 & 2 \\ -1 & 0 & 1 \end{bmatrix} = \begin{bmatrix} 1 \\ 2 \\ 1 \end{bmatrix} \begin{bmatrix} -1 & 0 & 1 \end{bmatrix} \quad (1)$$

$$Sy = \begin{bmatrix} -1 & -2 & -1 \\ 0 & 0 & 0 \\ 1 & 2 & 1 \end{bmatrix} = \begin{bmatrix} -1 \\ 0 \\ 1 \end{bmatrix} \begin{bmatrix} 1 & 2 & 1 \end{bmatrix} \quad (2)$$

The straight-forward approach to image convolutions used in related research is an adder tree [2, 4]. Improvements on this method have rightly focused on reducing the number of multiplications because this is often the cause of bottlenecks. However, multiplications can be avoided altogether by using the Sobel filters. Multiplying by 0 and 1 are not multiplications at all, and, in digital electronics, multiplication by 2 is a 1-bit shift left that requires a single clock cycle to perform.

Separation of the Sobel operators into row and column vectors is shown in (1) and (2), which is not possible with all edge detectors. Doing this does not affect the results of the image convolution. In fact, the result is independent of the order in which they are performed. With the separated kernels, 2D convolutions become two 1D convolutions. Observe from Fig. 1 that as the kernel shifts over the image along the arrow, the result of each column-vector convolution is used three times. Therefore, these convolutions need only be determined once and the results cached for subsequent use. With caching, there are four operands involved in each convolution, multiplication by two, subtraction and addition. The Sobel detector is used for the new architecture proposed in this paper.

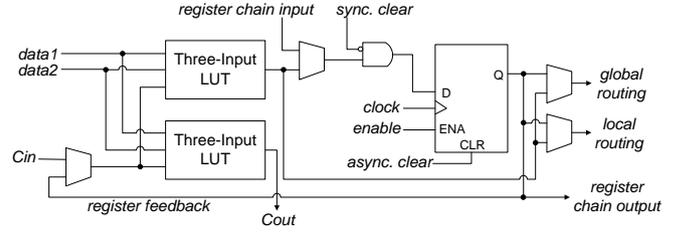

Fig. 2. Simplified logic element, in arithmatic mode, of an Altera Cyclone FPGA showing internal look-up tables and fast carry chain, adapted from [5].

*B. Field Progammable Gate Arrays*

Improvements in edge detection on FPGAs have taken cues from ASIC design but have not taken into account the specialised internal structure of FPGAs. FPGAs consist of a structured grid of interconnected logic elements (LE), such as an Altera Cyclone FPGA (Fig. 2). Combinational functions are synthesised as look-up tables (LUT) that are connected with multiplexed busses. This is the fundamental difference to ASIC design. No matter the intended architecture of the design, it will be synthesised using the LUT structure of the FPGA. The result is functionally identical to an ASIC design but can potentially have vastly different performance characteristics because of very different critical paths.

The LUT structure is designed for full-adders, with the fast carry-out chains connecting directly to adjacent logic elements for rapid carry propagation. A common exploit in FGPA design is to use these circuits instead as 3:2 compressors, which accepts three input operands and compresses them into two output operands. The result is that, after compression, only one addition is required to add three operands instead of two. The subtle difference between this behaviour and that of a full-adder is the carry-propagation. In a full-adder, carry bits must propagate the entire length of the add chain but a 3:2 compressor requires no logic propagation.

The numbers of logic elements in an FPGA are the largest dictator of its cost [6]. Power dissipation is also roughly proportional to the number of active logic elements in a design, because this defines the number of switching transistors. Methods such as pipelining [3, 7] that have traditionally been used in ASICs to increase throughput can greatly increase the power dissipation of an FPGA. To reduce the system cost and power dissipation of FPGA edge detection, the new architecture proposed here uses 3:2 compressors to better use the FPGAs resources and decrease overall usage. High-speed designs can reduce power dissipation of the overall system by freeing computation time for other tasks on the UAV (or equivalent) system that are also required in each control loop, so the FPGA can be clocked at a lower rate. The fast carry chains that connect adjacent logic elements are used in the new design to increase speed without increasing the area requirements of the design.

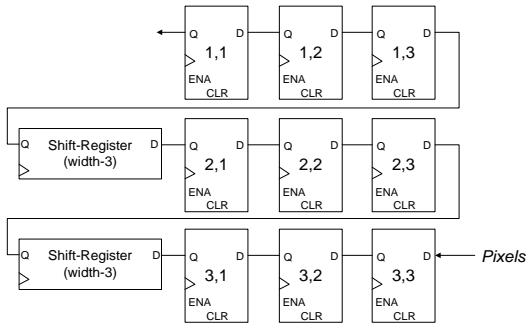

Fig. 3. Pixel cache that stores two image lines at a time and enables access to the data needed in each convolution. The pipeline takes pixels row-by-row like it would be presented in a video-feed.

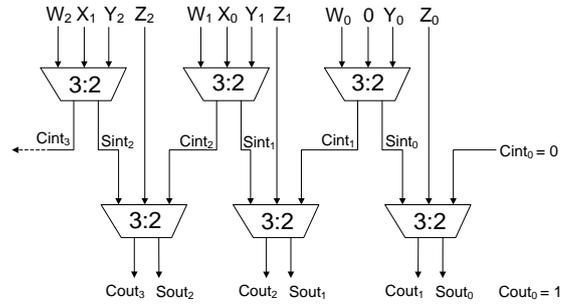

Fig. 4. Three least significant bits in a 4:2 compressor formed from two 3:2 compressors. The first layer performs multiplication and the second layer performs subtraction.

## III. FPGA Architecture

Here, the 3-input LUTs of an Altera Cyclone IV FPGA configured in arithmetic mode is considered. This type of configuration is typical of lower-cost FPGAs.

Convolution with the Sobel operators requires multiplication by two, subtraction and addition of four operands. Typical compressors function with only the addition of two's complement numbers so negation and multiplication must be performed prior to compression. Each operation has an instructional overhead that slows down operation. The new compressor outlined here avoids this by incorporating a subtraction and multiplication into a 4:2 compressor.

The two outputs from a compressor are added together to determine edge intensity. Ripple-carry addition is generally the fastest on FPGA, but Zicari and Perri [8] showed these can be sped up with some custom look-ahead logic. The new design uses their look-ahead adder to add the two outputs from the 4:2 compressor.

### A. Pixel Caching

Convolutions require some way of accessing the relevant pixel information for each calculation. The straightforward approach is to store the entire image in FPGA memory but this is a waste of resources and can create significant overheads from multiple memory accesses. A more area-efficient design (Fig. 3) caches only the image rows immediately needed by the convolution. The 3×3 grid of registers presents the relevant pixel data for each calculation and the shift registers buffer the rest of the row. Buffering the rows means that there is only one pixel entering the buffer each clock cycle. This reduces the dependency on memory access speeds and makes the design useful for receiving data directly from an image sensor. It is important to note that if the FPGA is to receive image data directly from a video-feed, there must be some mechanism for the cache pipeline to be primed, stalled and flushed [7]. This is necessary because of the horizontal blanking between lines and vertical blanking between frames that are present in video data.

Registers are drawn back-to-front from conventional orientation in Fig. 3 to illustrate the one-to-one mapping to the Sobel operators. Enable (ENA) and clear (CLR) inputs control the pipeline.

### B. Two-level 4:2 compressor

It is best to implement a 4:2 compressor using two levels of 3:2 compressors (Fig. 4), because it fits better with 3-input LUTs. Two types of compressors are required to meet the requirements of the separated Sobel filters. The first level of compressors combine the *W*, *X* and *Y* operands, multiplying Y by two, to interim sum and carry bits. The Boolean equations for the interim bits are shown in (3) and (4). Note the offset of the *Y* operator. The least significant *Y*-bit is set to zero. These are referred to as *P2PP:PP* compressors, where *2P* symbolises the doubling of a positive value.

$$Carry_{int,i+1} = (X_i \cdot Y_{i-1}) + (X_i \cdot Z_i) + (Y_{i-1} \cdot Z_i) \quad (3)$$

$$Sum_{int,i} = X_i \oplus Y_{i-1} \oplus Z_i \quad (4)$$

The second layer of compressors combines the interim bits and subtraction of the *Z* operand to complete the 4:2 compression. These are referred to as *PPN:PP* compressors, where *N* symbolises a number to be subtracted. The Boolean equations are shown in (5) and (6). Note that there is no equation for the least significant carry bit. Usually this is set to zero but the new compressor requires it to be 1. This requirement comes about from the subtraction stage.

$$Carry_{out,i+1} = \overline{Y_i} + X_i \cdot Y_i \cdot Z_i \quad (5)$$

$$Sum_{out,i} = \overline{X_i \oplus Y_i \oplus Z_i} \quad (6)$$

To function with the separated Sobel filters, some interim results must be cached for later use. Position of the caching differs slightly for the two directions of Sobel filters (Fig. 5).

### C. Faster Addition

Addition on FPGAs has been optimised for ripple-carry adders, which can be seen by the fast carry chains connecting logic elements. Many different adder designs have been proven to increase speed of ASICs but these do not translate well to FPGA. In general, propagation-type additions are the fastest attainable on an FPGA. Non-conventional numbering systems, such as redundant number, have been gaining traction in both ASIC and FPGA design, partly because they can eliminate carry propagation altogether. However, for the small bit lengths of most image data (usually 8-bits) redundant number adders still do not outperform conventional ripple-propagate adders. Kamp et al. [9] showed that redundant adders only begin to outperform them for bit lengths greater than 24-bits.

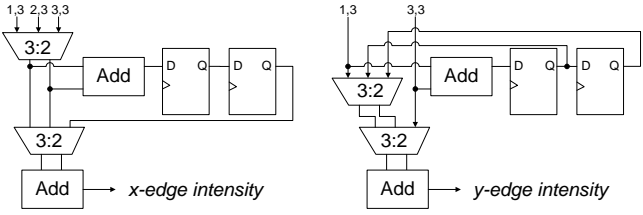

Fig. 5. Completed block diagram of new design showing caching and compression.

$$C_i = G_i + P_i \cdot C_{i-1} \quad (7)$$

$$G_i = (A_{i-1} \cdot B_{i-1}) + (A_{i-1} \oplus B_{i-1}) \cdot (A_{i-2} \cdot B_{i-2}) \quad (8)$$

$$P_i = (A_{i-1} \oplus B_{i-1}) \cdot (A_{i-2} \oplus B_{i-2}) \quad (9)$$

Despite the fast carry-chains on FPGAs, ripple-propagate adders are still limited by the time for carry-bits to travel the length of the adder chain. Zicari and Perri proved that breaking the propagation chain in two could increase performance. A generic n-bit adder is split into two n/2-bit adders with custom carry-prediction logic to remove the carry dependency of the most significant adder. The custom logic they developed is shown in (7-9), which allows the two half-length adders to be performed in parallel.

*D. FPGA Implementation*

FPGAs contain several dedicated circuits for performing multiple tasks at high data rates. One such task is fast shift registers that is used to efficiently implement pixel caching. Use of the dedicated DSP blocks can be forced with use of the ALTSHIFT_TAPS primitive [10].

To implement the 4:2 compressor, the two 3:2 compressors are split over two adjacent logic elements (LE), as shown in Fig. 6. The fast carry chain allows rapid propagation of the intermediate carry signal between LEs. For the FPGA to behave this way, the carry chain has to be initialised with the previous LE. The chain must also be terminated by another LE, which is not shown here. Chain termination can occupy the chain initialisation LE of the next compressor and therefore each compressor requires three LEs. For the *x*-direction convolution, another LE is needed between the compressors to perform the addition required at the start of the cache pipeline. To force this resource usage requires use of the low-level CARRY_SUM primitive [11]. However, the design was tested without the primitives and the synthesiser set to automatically create carry-chains, giving the same performance as with the primitives. This shows that the synthesiser efficiently creates carry chains for compressors and uses dedicated RAM blocks for shift registers.

The custom carry look-ahead logic in (7-9) depends on the same four operand bits. When a logic element is configured in normal mode, the LUTs have four inputs. Therefore, the look-ahead logic requires only one logic element to be performed. Breaking the carry-propagation distance also allows the adders to occupy a smaller section in each column of LEs and maximise carry propagation speed. When adder chains do not have enough room on a single column (columns have 16 LEs, some of which may be used by other functions), slow global routing must be used to connect the next column.

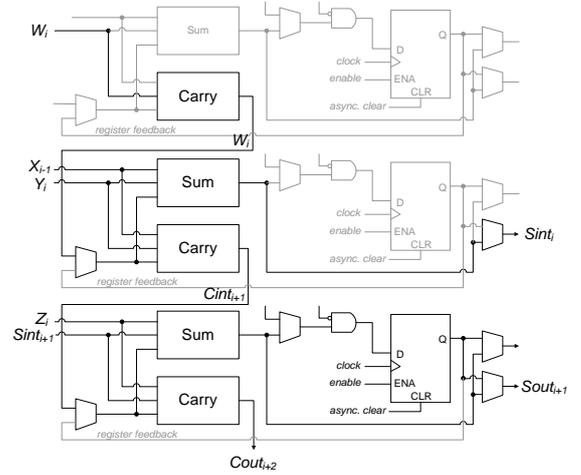

Fig. 6. 4:2 compressor inside an Altera FPGA. The topmost LE intitiates the carry chain, the second LE is a *P2PP:PP* compressor, and the bottom LE is a *PPN:PP* compressor. Logic paths are shown in black.

IV. RESULTS

Three tests were run to assess the performance of the new edge detection design: *1)* A visual examination of the detected edges to confirm that algorithm is behaving as expected; *2)* The speed of the design was tested with Altera's TimeQuest analysis package; *3)* The results of the FPGA fitter in the Quartus II compiler give the quantity and types of FGPA resources used by the design. Maximum operating frequency and FPGA resources are compared with Alghurav and Al-Rawi's [4] project on FPGA Sobel edge detection, Duncan's [12] Gumstik DSP Sobel implementation and Neoh and Hazanchuk's [3] FPGA Canny detector. The resource usage of these reference designs is difficult to compare because of differences in FPGA technology and compilers. To get a better picture of the improvements, four designs were implemented on FPGA based on different parts of the design proposed in this paper. *1)* The "Adder Tree" design uses the caching structure from III-A with an un-optimised adder tree. *2)* The "Separated" design adds caching and pipeline to the adder tree design, which is similar to Neoh and Hazanchuk's design. *3)* The "Compressor" design uses the 4:2 compressor outlined in Section III-B with a standard ripple-carry adder. *4)* The "Look-ahead Compressor" design is the full design outlined here.

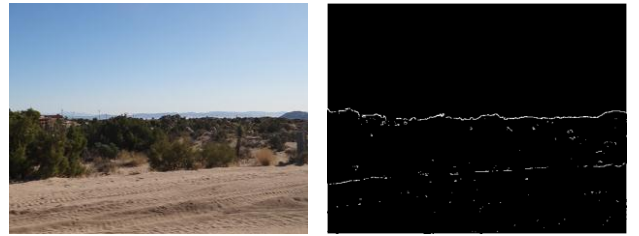

Fig. 7. Visual results of the new edge detection architecture. Left, image input. Right, image output showing detected edges in white.

Fig. 7 shows the visual edge detection results. A strong concentration of edges, shown in white, corresponds to the location of the horizontal, which was expected and verifies the correct functionality of the detector.

TABLE I.  MAXIMUM CLOCK FREQUENCY OF EDGE DETECTORS

| Edge Detector | Maximum Frequency (MHz) |
|---|---|
| Gumstik DSP [12] | 43.02 |
| Sobel reference design [4] | 169.12 |
| *Adder Tree* | 172.83 |
| *Separated* | 235.18 |
| Canny reference design [3] | 264.00 |
| *Compressor* | 321.89 |
| *Look-Ahead Compressor* | 338.41 |

a. Italics show the designs that were created as part of this paper.

Table 1 summarises the maximum operating frequency of the designs in megahertz. Table 2 summarises the resource usage of the designs. The number of LEs is broken into the number of combinational functions implemented on LE LUTs and LE registers. The new design proposed in this paper is 28% faster than the best-performing Canny design and uses 84% fewer FPGA logic elements. Dedicated registers show the number of registers used in caching is decreased by the compressor. This value will scale with the size of the image, which was 512×512 by 8-bits for testing.

TABLE II.  RESOURCE USAGE OF EDGE DETECTORS

| Edge Detector | Total LEs | LUTs | Logic Registers | Dedicated Registers |
|---|---|---|---|---|
| Gumstik DSP [12] | NA | NA | NA | NA |
| Sobel reference design [4] | 2543 | NA | NA | NA |
| *Adder Tree* | 258 | 223 | 115 | 8,128 |
| *Separated* | 272 | 180 | 190 | 8,160 |
| Canny reference design [3] | 1530 | NA | NA | NA |
| *Compressor* | 243 | 231 | 100 | 8,144 |
| *Look-Ahead Compressor* | 260 | 240 | 105 | 8,158 |

b. Italics show the designs that were created as part of this paper.

Although the number of combination functions increases by 60 when implementing the proposed design (in reference to the *Separated* design), the total LEs decrease by 12. This illustrates that the new design is better suited to FPGA implementation and will require fewer resources than past designs.

## V. CONCLUSIONS

To decrease system cost and power dissipation in edge detection, a new FPGA architecture was proposed to exploit the specialised internal structure of FPGAs. Image buffering is minimised by caching two rows of an image. The 4-input LUTs of an Altera FPGA are repurposed as 4:2 compressors, constructed from two custom-designed 3:2 compressors. After compression, only one addition operation is required to find edge intensity. The new design therefore decreases computational complexity from 17 operations, required by full 3×3 convolutions, to 1 operation. Ripple-carry addition is sped up by breaking carry-propagation with look-ahead logic.

Although the design was specifically intended for low-cost FGPAs with 4-input LUTs, the new design would still function on FPGAs with different LUT structures. Xilinx's Virtex family of FPGAs use 6-input LUTs that require more silicon but allow larger functions to be implemented without intermediate steps. Xilinx FPGAs also feature several features that Altera Cyclone FPGAs do not, such as dedicated logic gates. These two features could further decrease latency and overall area requirements of the new design.

Improved behaviour was confirmed with experimental results that showed a 28% increase in speed and 4.4% decrease in resources over current edge detectors. The new design is ready for integration into UAVs to decrease system cost and increase flight-time by dissipating less power.

REFERENCES

[1] J. Stowers, A. Bainbridge-Smith and M. Hayes, "Beyond Optical Flow - Biomimetic UAV Altitude Control Using Horizontal Edge Information," in *Proceedings of the 5th International Conference on Automation, Robotics and Applications*, Wellington, New Zealand, 2011.

[2] M. Juneja and P. S. Sandhu, "Performance Evaluation of Edge Detection Techniques for Images in Spatial Domain," *International Journal of Computer Theory and Engineering,* vol. 1, no. 5, pp. 614-621, 2009.

[3] H. S. Neoh and A. Hazanchuk, "Adaptive Edge Detection for Real-Time Video Processing using FPGAs," *Global Signal Processing,* 2004.

[4] D. Alghurair and S. S. Al-Rawis, "Design of Sobel operator using Field Programmable Gate Arrays," in *International Conference on Technological Advances in Electrical, Electronics and Computer Engineering (TAEECE)*, 2013.

[5] Altera Corperation, "Logic Elements and Logic Array Blocks in Cyclone IV Devices," in *Cyclone IV Device Handbook*, vol. 1, 2009.

[6] Altera Corperation, "Cyclone IV FPGA Device Family Overview," in *Cyclone IV Device Handbook*, vol. 1, 2013.

[7] C. T. Johnston, K. T. Gribbon and D. G. Bailey, "Implementing Image Processing Algorithms on FPGAs," in *Proceedings of the Eleventh Electronics New Zealand Conference (ENZCon'04)*, Palmerston North, New Zealand, 2004.

[8] P. Zicari and S. Perri, "A Fast Carry Chain Adder for Virtex-5 FPGAs," in *MELECON 2010 - 15th IEEE Mediterranean Electrotechnical Conference*, Valletta, 2010.

[9] W. Kamp, A. Bainbridge-Smith and M. Hayes, "Efficient Implementation of Fast Redundant Number for Long Word-lengths in FPGAs," in *International Conference of Field-Programmable Technology*, Syndney, NSW, 2009.

[10] Altera Corperation, "RAM-Based Shift Register (ALTSHIFT_TAPS) Megafunction," in *Altera Primitives Megafunction User Guide*, 2013.

[11] Altera Corperation, "Designing with Low-Level Primitives," in *Altera Primitives User Guide*, 3 ed., 2007.

[12] S. Duncan, "DSP Acceleration for Real Computer Vision on Embedded Platforms," 2012.